\documentclass[onecolumn, floatfix, preprintnumbers, eqsecnum, letterpaper, superscriptaddress, nofootinbib]{revtex4}

\usepackage{graphicx}
\usepackage{microtype}
\usepackage{amsmath}
\usepackage{amssymb}
\usepackage{subfigure}
\usepackage{hyperref}
\usepackage{url}
\usepackage{xcolor}
\usepackage{color}
\usepackage{mathrsfs}
\usepackage{calrsfs}
\usepackage{amsfonts}
\usepackage{eufrak}
\usepackage{tabularx}
\usepackage{eucal}
\usepackage{latexsym}
\usepackage{ragged2e}
\usepackage{epsfig}
\usepackage{textcomp}
\usepackage{caption}
\usepackage{subfigure}
\usepackage{marvosym}
\usepackage{ifsym}
\usepackage{lipsum}
\usepackage{appendix}
\usepackage{bm}
\usepackage{mathrsfs}

\DeclareCaptionJustification{justified}{\leftskip=0pt \rightskip=0pt \parfillskip=0pt plus 1fil}
\definecolor{vividviolet}{rgb}{0.62, 0.0, 1.0}
\definecolor{amaranth}{rgb}{0.9, 0.17, 0.31}
\definecolor{palatinateblue}{rgb}{0.15, 0.23, 0.89}
\definecolor{brightpink}{rgb}{1.0, 0.0, 0.5}
\definecolor{cornflowerblue}{rgb}{0.39, 0.58, 0.93}
\definecolor{deepcarminepink}{rgb}{0.94, 0.19, 0.22}
\definecolor{radicalred}{rgb}{1.0, 0.21, 0.37}

\hypersetup{ linktoc=all,
    colorlinks, linkcolor={palatinateblue},
    citecolor={brightpink}, urlcolor={amaranth}
}

\graphicspath{{Images/}}

\graphicspath{{Images/}}

\def\@fnsymbol#1{\ensuremath{\ifcase#1\or \ddagger \or  $\textleaf$  \or \dagger
\else\@ctrerr\fi}}%

\makeatother

\def\sideremark#1{\ifvmode\leavevmode\fi\vadjust{\vbox to0pt{\vss
 \hbox to 0pt{\hskip\hsize\hskip1em
 \vbox{\hsize1.3cm\tiny\raggedright\pretolerance10000
 \noindent #1\hfill}\hss}\vbox to8pt{\vfil}\vss}}}%

\def\beq{\begin{equation}}
\def\eeq{\end{equation}}

\newcommand{\od}{\mathrm{d}}

\begin{document}

\title{Equation of State and Joule-Thomson Expansion for the FRW Universe in the Brane World Scenario}

\author{Shi-Bei Kong}
\email{kongshibei@nuaa.edu.cn}
\affiliation{College of Physics, Nanjing University of Aeronautics and Astronautics, Nanjing, 211106, China}

\author{Haximjan Abdusattar}
\email{axim@nuaa.edu.cn}
\affiliation{College of Physics, Nanjing University of Aeronautics and Astronautics, Nanjing, 211106, China}

\author{Hongsheng Zhang}
\email{sps$_$zhanghs@ujn.edu.cn}
\affiliation{School of Physics and Technology, University of Jinan, 336 West Road of Nan Xinzhuang, Jinan, Shandong 250022, China}
\affiliation{Key Laboratory of Theoretical Physics, Institute of Theoretical Physics, Chinese Academy of Sciences, Beijing 100190, China}

\author{Ya-Peng Hu }
\email{huyp@nuaa.edu.cn}
\affiliation{College of Physics, Nanjing University of Aeronautics and Astronautics, Nanjing, 211106, China}
\affiliation{Key Laboratory of Aerospace Information Materials and Physics (NUAA), MIIT, Nanjing 211106, China}

\begin{abstract}

We study the thermodynamic properties of the Friedmann-Robertson-Walker (FRW) universe in the brane world scenario, concentrating on the Randall-Sundrum II model. From the first law of thermodynamics for the FRW universe, we find that the work density $W$ can be identified with the thermodynamic pressure $P$. We construct the equation of state $P$$=$$P(V, T)$ for the FRW universe in the brane world scenario, which does not show $P$-$V$ phase transition. We further study the Joule-Thomson expansion of the FRW universe, and derive the Joule-Thomson coefficient,
which has an inversion point that is affected by the brane tension. These results could provide new ways to test the brane world scenario
and extra dimension.

\bigskip

\end{abstract}

\maketitle

\section{Introduction}

There are rich thermodynamic information hidden in the Friedmann's equations for the FRW universe
\cite{Cai:2005ra,Akbar:2006er,Akbar:2006kj,Sheykhi:2007gi,Cai:2008ys}.
From the Friedmann's equations or field equations, one can get the unified first law \cite{Hayward:1997jp,Cai:2006rs}
for the FRW universe, which gives the first law of thermodynamics (Gibbs equation) after projecting on the apparent horizon of the FRW universe.
From the first law of thermodynamics, one can read out many thermodynamic quantities,
such as internal energy, and pairs of conjugate variables such as temperature/entropy, thermodynamic pressure/volume, etc.
Especially \cite{Abdusattar:2021wfv,Kong:2021dqd}, we found that the thermodynamic pressure can be defined from the work density
\begin{alignat}{1}
P:=-\frac{1}{2}h_{ab}T^{ab}, \label{tp}
\end{alignat}
and its conjugate variable is the thermodynamic volume $V=4\pi R_A^3/3$, where $R_A$ is the physical radius of the apparent horizon.
Inspired by the equation of state $P=P(V,T)$ and phase transitions for kinds of AdS black holes \cite{Dolan:2010ha,Kubiznak:2012wp,Cai:2013qga,Hu:2018qsy,Hu:2020pmr},
we have also established similar equations and studied phase transitions for the FRW universe \cite{Abdusattar:2021wfv,Kong:2021dqd}
as well as McVittie black hole \cite{Abdusattar:2022bpg}.
We found that the FRW universe has a  ``small-large" (or $P$-$V$) phase transition \cite{Kong:2021dqd} and the McVittie spacetime
has a Hawking-Page-like phase transition  \cite{Abdusattar:2022bpg}.

Inspired by previous findings, we would like to further investigate the thermodynamics of the FRW universe but adopt other gravitational theories. In previous studies \cite{Abdusattar:2021wfv,Kong:2021dqd}, we did not consider field equations in other modified theories of gravity, which may modify the Friedmann's equations and thus change the thermodynamical behaviors of the FRW universe.
Therefore, in this work, we would like to investigate the thermodynamics of the FRW universe under other gravitational theories. Note that, the brane world scenario
\cite{Rubakov:1983bb,Rubakov:1983bz,Bergshoeff:1987cm,Dai:1989ua,Horowitz:1991cd,Horava:1995qa,Polchinski:1996na,Aharony:1997ju,Arkani-Hamed:1998jmv,Dienes:1998vh,Antoniadis:1998ig,Randall:1999ee,Randall:1999vf,Hawking:2000kj,Lust:2003ky,Verlinde:2005jr,Liu:2011wi}
is an important theory beyond standard model and general relativity (GR).
If extra dimensions exist, we are naturally led to this fascinating scenario.
Extra dimensions were first introduced in the Kaluza-Klein theory,
where the extra dimensions are compact and very small. However, in the brane world scenario, extra dimensions may be very large,
which can help to solve the hierarchy problem in the standard model of particle physics \cite{Langlois:2002bb,Meng:2002jz}.
In this scenario, our Universe is treated as a 3-brane (object with 3 spatial dimensions) embedded in a higher dimensional spacetime (bulk). Standard model fields are constrained (localized) on the brane, but gravitons can propagate in all dimensions \cite{Arkani-Hamed:1998jmv,Shiromizu:1999wj}. This scenario offers an alternative explanation of the acceleration \cite{SupernovaSearchTeam:1998fmf,SupernovaCosmologyProject:1998vns}
of our Universe without introducing cosmological constant \cite{Peebles:2002gy,Tye:2017upp} and quintessence-like matter. In the brane world scenario, there is a Randall-Sundrum II (RSII) model \cite{Randall:1999vf}, which is relatively simple but captures the very essence of this scenario. In the RSII model, the Friedmann's equations for the FRW universe are different from those in GR, which implies that the thermodynamics of the FRW universe in this model may be different from that in GR. Therefore, in this paper, we make a detailed thermodynamic analysis of the FRW universe in the brane world scenario. Especially, we study the equation of state and Joule-Thomson expansion of the FRW universe in the brane world scenario, and compare the results with those in GR \footnote{GR is a limit case of the brane world scenario.}.
Remarkably, we find that the equation of state of the FRW universe in the brane world scenario is very different from that in GR. Moreover, inversion points of JT expansion process exist and are affected by the brane tension or the extra dimension. These results may shed light on new ways to test the brane world scenario or extra dimension.

This paper is organized in the following way.
In Sec.II, we introduce the RSII model and the FRW universe in the brane world scenario.
In Sec.III, we investigate the first law of thermodynamics for the FRW universe in the brane world scenario.
In Sec.IV, we construct the equation of state for the FRW universe in the brane world scenario.
In Sec.V, we investigate the Joule-Thomson expansion of the FRW universe in brane world scenario.
In the last section, we give conclusions and make some discussions.
In this paper, we choose units $c=\hbar=k_B=1$.

\section{the RSII Model and the FRW Universe in the Brane World Scenario}

In this section, we make a brief introduction of the RSII model and the FRW universe in the brane world scenario.

In the RSII model, the FRW universe is treated as a 3-brane embedded in a 5-dimensional (5d) bulk with $\mathbb{Z}_2$ symmetry.
Gravity in the 5d bulk is still general relativity
\begin{alignat}{1}
R_{\alpha\beta}-\frac{1}{2}g_{\alpha\beta}R=\kappa_5^2 T_{\alpha\beta},
\end{alignat}
but its reduction on the brane is different from GR \cite{Shiromizu:1999wj, Aliev:2004ds}:
\begin{alignat}{1}
R_{\mu\nu}-\frac{1}{2}g_{\mu\nu}R=-\Lambda_4 q_{\mu\nu}+8\pi G_4\tau_{\mu\nu}+\kappa_5^4\pi_{\mu\nu}-E_{\mu\nu}, \label{feb}
\end{alignat}
where
\footnote{$\Lambda_4$ is the effective cosmological constant on the brane, $q_{\mu\nu}$ is the induced metric on the brane,
$n^{\mu}$ is the unit normal vector of the brane, $G_4$ is the 4d Newton constant,
$\tau_{\mu\nu}$ is the energy-momentum tensor of ordinary matter on the brane,
$\kappa_5$ is the 5d gravity coupling constant, $E_{\mu\nu}$ is the electric part of the 5d Weyl tensor,
$\lambda$ is the brane tension, and $\Lambda_5$ is the 5d cosmological constant.}
\begin{alignat}{1}
\Lambda_4=&\frac{1}{2}\kappa_5^2\left(\Lambda_5+\frac{1}{6}\lambda^2\kappa_5^2\right),
\quad q_{\mu\nu}=g_{\mu\nu}-n_{\mu}n_{\nu}, \quad G_4=\frac{\lambda}{48\pi}\kappa_5^4,
\nonumber \\
\pi_{\mu\nu}=&-\frac{1}{4}\tau_{\mu\alpha}\tau^{\alpha}_{~\nu}+\frac{1}{12}\tau\tau_{\mu\nu}
+\frac{1}{8}q_{\mu\nu}\tau_{\alpha\beta}\tau^{\alpha\beta}-\frac{1}{24}q_{\mu\nu}\tau^2. \label{pi}
\end{alignat}
In the following, we choose the 5d AdS space as the bulk, so the electric part $E_{\mu\nu}$ of the 5d Weyl tensor can be ignored.
We also set $\Lambda_4$ to be zero
\footnote{The effective cosmological constant $\Lambda_4$ on the brane can take arbitrary value
\cite{Shiromizu:1999wj}, so for convenience, we set it to be zero as in \cite{Cai:2006pa}.},
so there is only one additional term $\kappa_5^4\pi_{\mu\nu}$ compared with 4d Einstein field equation.
Therefore, the field equation (\ref{feb}) can be written as
\begin{alignat}{1}
R_{\mu\nu}-\frac{1}{2}g_{\mu\nu}R=8\pi G_4(\tau_{\mu\nu}+\overset{e}{\tau}_{\mu\nu}), \label{fe}
\end{alignat}
where
\begin{alignat}{1}
\overset{e}{\tau}_{\mu\nu}\equiv\frac{\kappa_5^4}{8\pi G_4}\pi_{\mu\nu}=\frac{6\pi_{\mu\nu}}{\lambda}.
\end{alignat}
Taking the $\lambda\rightarrow\infty$ limit of (\ref{fe}), one can get the Einstein field equation, i.e. recover general relativity.

In order to get the modified Friedmann's equations from (\ref{fe}), one should insert the line-element of the FRW universe
and the energy momentum tensor $\tau_{\mu\nu}$ of ordinary matter.
The line-element of the FRW universe can be written as
\begin{alignat}{1}
\od s^2=q_{\mu\nu}\od x^{\mu}\od x^{\nu}=&-\od t^2+a(t)^2\left[\frac{1}{1-k r^2}\od r^2
+r^2(\od\theta^2+\sin^2\theta\od\varphi^2)\right], \label{metric}
\end{alignat}
and the energy momentum of ordinary matter can be written in the form of a perfect fluid
\begin{alignat}{1}
\tau_{\mu\nu}=(\rho+p)U_{\mu}U_{\nu}+pq_{\mu\nu},
\end{alignat}
where $\rho$ is the energy density, $p$ is the pressure and $U^{\mu}$ is the normalized 4-velocity, i.e. $U_{\mu}U^{\mu}=-1$.
Combined with (\ref{pi}), one can also get
\begin{alignat}{1}
\pi_{\mu\nu}=\frac{1}{6}\rho\left[(\rho+p)U_{\mu}U_{\nu}+\left(\frac{1}{2}\rho+p\right)q_{\mu\nu}\right].
\end{alignat}

Thus one can get the Friedmann's equations in the RSII model \cite{Cai:2006pa}
\begin{alignat}{1}
H^2+\frac{k}{a^2}=&\frac{8\pi G_4}{3}(\rho+\rho_e), \label{fe1}
\\
-\left(\dot{H}-\frac{k}{a^2}\right)=&4\pi G_4(\rho+\rho_e+p+p_e), \label{fe2}
\end{alignat}
where $\rho$ and $p$ of the ordinary matter satisfy the continuity equation
\begin{alignat}{1}
\dot{\rho}+3H(\rho+p)=0, \label{continuity}
\end{alignat}
and
\begin{alignat}{1}
\rho_e\equiv&\frac{\rho^2}{2\lambda}, \quad p_e\equiv\frac{\rho}{\lambda}\left(\frac{\rho}{2}+p\right) \label{rhoe}
\end{alignat}
are the effective energy density and pressure reduced from the extra dimension.
In the first modified Friedmann's equation (\ref{fe1}), the $\rho_e$ term is a high energy correction to the standard Friedmann's equation because it dominates in the high energy regime $\rho>\lambda$ (such as in the early period of our Universe)
and is negligible in the low energy regime $\rho\ll\lambda$ \cite{Binetruy:1999ut,Shiromizu:1999wj,Langlois:2002bb}.

From the Friedmann's equations (\ref{fe1}) and (\ref{fe2}), one can get the first law of thermodynamic
for the FRW universe in the brane world scenario. The first law of thermodynamics is established at the apparent horizon instead of
event horizon of the FRW universe, because the apparent horizon is a more practical notion than event horizon for the FRW universe
as well as dynamical black holes. Event horizon is defined in asymptotically flat space-times \cite{Hayward:1993wb},
and requires the knowledge of the global property rather than the local property of the spacetime \cite{Bardeen:1973gs},
i.e. it can not be located by ordinary observers \cite{Hayward:2004dv, Hayward:2004fz}.
However, the FRW universe is a dynamical (non-stationary) spacetime and not asymptotically flat \cite{Hayward:2008jq},
so the notion of event horizon is not applicable to it. The existence of the apparent horizon
depends on the local property of the spacetime, so it is more practical and can be used for the FRW universe
\footnote{Apparent horizon can be used in various dynamical spacetimes, from dynamical (such as evaporating) black hole, white hole,
traversable wormhole to cosmological models (such as the FRW universe), etc \cite{Hayward:2004dv, Hayward:2004fz}.
The longstanding black hole information paradox may also have a simple resolution \cite{Hayward:2008jq} if the apparent horizon is used
instead of the event horizon.}.

The apparent horizon of the FRW universe is defined as \cite{Bak:1999hd, Cai:2005ra}
\begin{alignat}{1}
h^{ab}\partial_a R\partial_b R=0,
\end{alignat}
where $a,b=0,1$ and $R\equiv a(t)r$ is the physical radius
\footnote{The line-element and some related concepts in the physical coordinate system of the FRW universe
is presented in Appendix A.}.
Combined with the line-element (\ref{metric}), one can get the location of the apparent horizon
\begin{alignat}{1}
R_A=\left(H^2+\frac{k}{a(t)^2}\right)^{-1/2}, \label{ra}
\end{alignat}
where $H:=\dot{a}(t)/a(t)$ is the Hubble parameter.
From (\ref{ra}), one can also get a very useful relation
\begin{alignat}{1}
\dot{R}_A=-HR_A^3\left(\dot{H}-\frac{k}{a(t)^2}\right). \label{dotra}
\end{alignat}

Associated with the apparent horizon, there is a surface gravity. It can be calculated from its definition
\footnote{For dynamical black hole, the definition (\ref{sg}) captures the information of the dynamical geometry,
such as evaporation rate, which may be useful in the resolution of the information puzzle \cite{Hayward:2008jq}.}
\cite{Cai:2006pa,Hayward:2008jq}
\begin{alignat}{1}
\kappa:=\frac{1}{2\sqrt{-h}}\partial_a(\sqrt{-h}h^{ab}\partial_b R), \label{sg}
\end{alignat}
and the result is
\begin{alignat}{1}
\kappa=-\frac{1}{R_A}(1-\epsilon), \label{kappa}
\end{alignat}
where $\epsilon:=\dot{R}_A/(2HR_A)$.
The surface gravity (\ref{kappa}) can be also written by the Ricci scalar and apparent horizon of the FRW universe as
\begin{alignat}{1}
\kappa=-\frac{R_A R}{12}.
\end{alignat}
Similar to the Hawking temperature, one can define the Kodama-Hayward temperature from the surface gravity (\ref{kappa})
\begin{alignat}{1}
T:=\frac{|\kappa|}{2\pi}, \label{KHT}
\end{alignat}
where the absolute value of the surface gravity is taken to guarantee that the temperature is always positive.
In this paper, we treat $\epsilon$ as a small quantity, i.e. $|\epsilon|<1$, which means that the physical radius of the apparent horizon
changes slowly
\footnote{In this case, the FRW universe can be treated as a quasi-equilibrium thermodynamic system.}.
In this case, the surface gravity $\kappa$ is negative, and the temperature is
\begin{alignat}{1}
T=-\frac{\kappa}{2\pi}=\frac{1}{2\pi R_A}(1-\epsilon).
\end{alignat}

\section{First Law of Thermodynamics for the FRW Universe in the Brane World Scenario}

In this section, we investigate the first law of thermodynamics at the apparent horizon of the FRW universe from the unified first law
\cite{Hayward:1997jp, Hayward:1998ee, Cai:2006rs}.

The unified first law in the brane world scenario is obtained from the ``00" component of the field equation (\ref{fe})
\begin{alignat}{1}
\od M=A\Psi_{a}\od x^{a}+W\od V, \label{ufl}
\end{alignat}
where
\begin{alignat}{1}
M:=&\frac{R}{2G_4}(1-h^{ab}\partial_{a}R\partial_{b}R), \quad A:=4\pi R^2, \quad V:=\frac{4\pi R^3}{3}, \label{av}
\\
W:=&-\frac{1}{2}h_{ab}T^{ab}, \quad \Psi_{a}:=T_{a}^{~b}\partial_{b}R+W\partial_{a}R,\label{workdensity}
\end{alignat}
are the Misner-Sharp energy, area, volume, work density, and energy-supply vector respectively.
It should be noted that the above work density $W$ and energy supply vector $\Psi_a$ are defined from the ``total" energy momentum tensor
$T_{ab}\equiv\tau_{ab}+\overset{e}{\tau}_{ab}$.

Project the unified first law (\ref{ufl}) along the tangent vector $\xi=\partial_t-(1-2\epsilon)\partial_r$ \cite{Cai:2006pa}
of the apparent horizon, one can get the first law of thermodynamics
\footnote{It can be also written by abstract notation \cite{Wu:2009wp}
\begin{alignat}{1}
\xi^a\nabla_a (-M)=-\frac{\kappa}{8\pi G_4}\xi^a\nabla_a A-W\xi^a\nabla_a V
=T\xi^a\nabla_a S-W\xi^a\nabla_a V.  \nonumber
\end{alignat}}
\begin{alignat}{1}
\langle\od(-M),\xi\rangle=-\frac{\kappa}{8\pi G_4}\langle\od A, \xi\rangle-\langle W\od V,\xi\rangle
=T\langle\od S, \xi\rangle-\langle W\od V,\xi\rangle, \label{fl}
\end{alignat}
where $S=A/(4G_4)$ is the Bekenstein-Hawking entropy. Compared with the ``standard" first law
\footnote{Strictly speaking, it is the Gibbs equation \cite{Hayward:1997jp,Wu:2009wp}.} of thermodynamics $\delta U=T\delta S-P\delta V$,
one can see that the work density $W$ plays the role of the thermodynamic pressure $P$, so the following identification can be made
\begin{alignat}{1}
P\equiv W=-\frac{1}{2}h_{ab}T^{ab}. \label{tp1}
\end{alignat}

One can also define work density $\overset{m}{W}$ and energy supply vector $\overset{m}{\Psi}_a$ from the energy-momentum tensor $\tau_{\mu\nu}$
of ordinary matter, and get the second formalism of the unified first law \cite{Cai:2006rs}
\begin{alignat}{1}
\od M_{eff}=A\overset{m}{\Psi}_a\od x^a+\overset{m}{W}\od V, \label{uflo}
\end{alignat}
where
\begin{alignat}{1}
\overset{m}{W}:=-\frac{1}{2}h_{ab}\tau^{ab}, \quad
\overset{m}{\Psi}_a:=\tau_{a}^{~b}\partial_{b} R+\overset{m}{W}\partial_{a}R. \label{emse}
\end{alignat}
In this case, the effective Misner-Sharp energy is found to be $M_{eff}=\rho V$.

In this case, the first law of thermodynamics is obtained by the projection of the unified first law (\ref{uflo})
along the tangent vector $\xi$ of the apparent horizon
\begin{alignat}{1}
\langle\od(-M_{eff}),\xi\rangle=\langle T\od\overset{m}{S}, \xi\rangle-\langle\overset{m}{W}\od V,\xi\rangle, \label{flo}
\end{alignat}
where
\begin{alignat}{1}
\overset{m}{S}=\frac{\pi R_A}{G_4}\sqrt{R_A^2+l^2}-\frac{\pi l^2}{G_4}\log\left(\frac{R_A}{l}+\sqrt{1+\frac{R_A^2}{l^2}}\right),
\quad l=\frac{\kappa_5^2}{8\pi G_4}, \label{entropy}
\end{alignat}
is the entropy of ordinary matter.
One can see that in this formalism: $-M_{eff}$ plays the role of the internal energy $U$;
$\overset{m}{W}$ plays the role of the thermodynamic pressure $P$.
Therefore, one can make these identifications
\begin{alignat}{1}
U\equiv-M_{eff}, \label{ie}
\end{alignat}
and
\begin{alignat}{1}
P\equiv\overset{m}{W}. \label{tp2}
\end{alignat}

\section{Equation of State $P=P(V,T)$ for the FRW Universe in the Brane World Scenario}

In this section, we derive the equation of state $P=P(V,T)$ for the FRW universe in the RSII model of the brane world scenario.
We can use either $P\equiv W$ or $P\equiv\overset{m}{W}$ in this derivation, but the first case formally resembles that in GR,
which is not interesting, so we put it in Appendix B. In the following, we derive the equation of state for $P\equiv\overset{m}{W}$.

At first, we get the energy density $\rho$ from the Friedmann's equations (\ref{fe1}) and (\ref{rhoe}),
which has two solutions:
\begin{alignat}{1}
\rho_+=&\frac{\sqrt{4\pi^2\lambda^2R_A^2+3\pi\lambda}}{2\pi G_4 R_A}-\frac{\lambda}{G_4},  \label{rho}
\\
\rho_-=&-\frac{\sqrt{4\pi^2\lambda^2R_A^2+3\pi\lambda}}{2\pi G_4 R_A}-\frac{\lambda}{G_4}.
\end{alignat}
The first solution $\rho_+$ can reduce to the one in GR when the brane tension $\lambda$ approaches infinity,
but the second solution $\rho_-$ can not, so we drop this solution.
Apply the solution $\rho_+$ to the continuity equation (\ref{continuity}), one can get the pressure
\begin{alignat}{1}
p=\frac{\lambda\dot{R}_A}{2G_4HR_A^2\sqrt{4\pi^2\lambda^2R_A^2+3\pi\lambda}}
-\frac{\sqrt{4\pi^2\lambda^2R_A^2+3\pi\lambda}}{2\pi G_4R_A}+\frac{\lambda}{G_4}. \label{1.14}
\end{alignat}

Therefore, the work density of ordinary matter is
\begin{alignat}{1}
W_m:=-\frac{1}{2}h_{ab}\tau^{ab}=\frac{1}{2}(\rho-p)
=\frac{\sqrt{4\pi^2\lambda^2R_A^2+3\pi\lambda}}{2\pi G_4 R_A}
-\frac{\lambda\dot{R}_A}{4G_4HR_A^2\sqrt{4\pi^2\lambda^2R_A^2+3\pi\lambda}}-\frac{\lambda}{G_4},
\end{alignat}
combined with (\ref{kappa}), (\ref{KHT}), (\ref{tp2}) and $V=4\pi R_A^3/3$, one can get the equation of state
\begin{alignat}{1}
P(V,T)=&\frac{\lambda[\pi T (12\pi V)^{1/3}+2\pi\lambda(3V)^{2/3}+(4\pi)^{2/3}]}
{G_4(3V)^{1/3}\sqrt{4\pi^2\lambda^2(3V)^{2/3}+3\pi\lambda (4\pi)^{2/3}}}-\frac{\lambda}{G_4},
\end{alignat}
but it is more concise to use the radius of the apparent horizon
\begin{alignat}{1}
P(R_A,T)=\frac{\lambda(\pi T R_A+2\pi\lambda R_A^2+1)}{G_4R_A\sqrt{4\pi^2\lambda^2R_A^2+3\pi\lambda}}
-\frac{\lambda}{G_4}. \label{eoso}
\end{alignat}
This equation does not have a critical point, so it has no $P$-$V$ (or small-large) phase transition.
\footnote{If $\lambda R_A^2\gg 1$, one can make a virial expansion \cite{Dolan:2011xt} for the equation of state (\ref{eoso})
\begin{alignat}{1}
P=\frac{T}{2G_4R_A}+\frac{1}{8\pi G_4R_A^2}+\frac{3}{128\pi^2\lambda G_4R_A^4}
-\frac{3T}{16\pi\lambda G_4R_A^3}+\mathcal{O}\left[\left(\frac{1}{\lambda R_A^2}\right)^{2}\right].
\end{alignat}}

\begin{figure}[h]
\includegraphics[width=7.7cm]{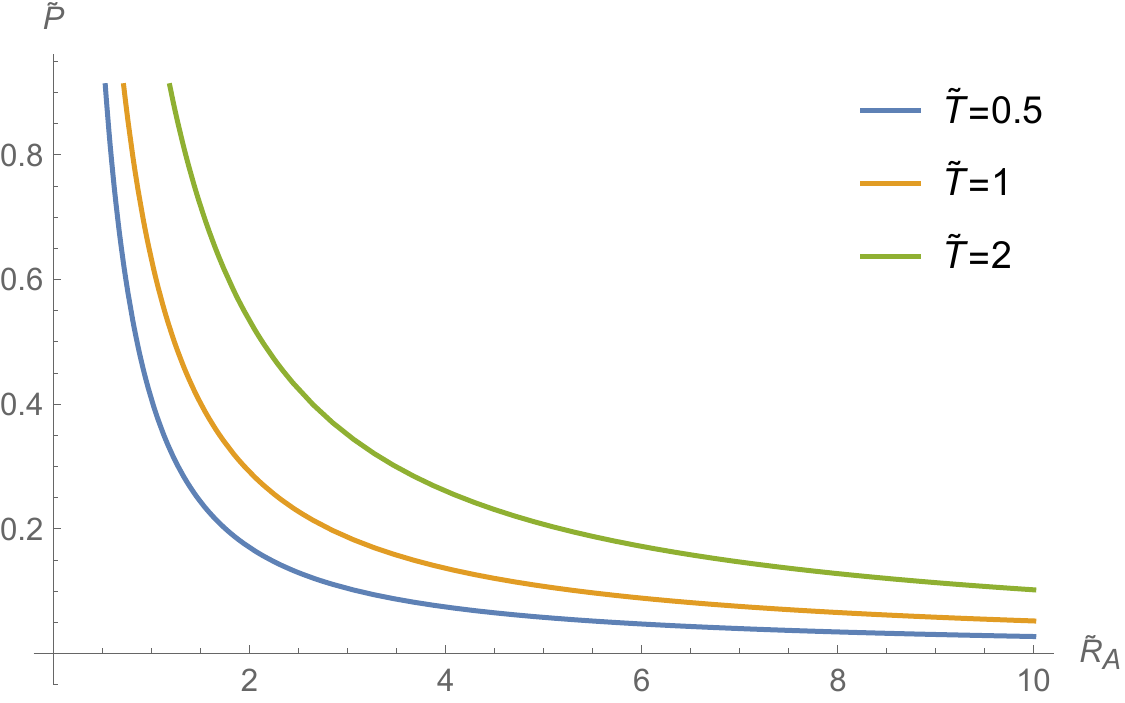}
\caption{Isothermal lines from the equation of state. These lines does not show $P$-$V$ phase transition.}\label{FIG.1}
\end{figure}

One can define dimensionless variables
\begin{alignat}{1}
\tilde{P}:=\frac{G_4}{\lambda}P, \quad \tilde{T}:=\frac{T}{\sqrt{\lambda}}, \quad \tilde{R}_A:=\sqrt{\lambda}R_A,
\end{alignat}
and rewrite the equation of state (\ref{eoso}) in a dimensionless form
\begin{alignat}{1}
\tilde{P}=\frac{\pi\tilde{T}\tilde{R}_A+2\pi R_A^2+1}{\tilde{R}_A\sqrt{3\pi+4\pi^2\tilde{R}_A^2}}-1,
\end{alignat}
which is more convenient for drawing diagrams, see FIG. \ref{FIG.1} for some isothermal lines from the above dimensionless equation.

\section{Joule-Thomson Expansion of the FRW Universe in the Brane World Scenario}

In this section, as an application of the equation of state (\ref{eoso}), we discuss the Joule-Thomson (JT)
expansion of the FRW universe in the brane world scenario.

In the JT expansion, the enthalpy of the system does not change, but the thermodynamic pressure keeps changing,
which results to the change of the temperature. The JT coefficient is defined to describe the change of temperature
with the thermodynamic pressure
\begin{alignat}{1}
\mu:=\left(\frac{\partial T}{\partial P}\right)_H, \label{mu1}
\end{alignat}
and the inversion point in the JT expansion has a zero JT coefficient.
Recent years, the JT coefficient has been calculated for some AdS black holes,
e.g. \cite{Okcu:2016tgt,Mo:2018rgq,Cisterna:2018jqg}, and the FRW universe in Einstein gravity \cite{Abdusattar:2021wfv}.

In the following, we derive the JT coefficient for the FRW universe in the brane world scenario.
At first, we get the enthalpy from (\ref{ie}) and (\ref{tp2})
\begin{alignat}{1}
H:=U+PV=-M_{eff}+PV=-\rho V+PV,
\end{alignat}
which should be fixed in the JT expansion.
From (\ref{rho}) and $V=4\pi R_A^3/3$, one can express $P$ as a function of $R_A$:
\begin{alignat}{1}
P(R_A)=\frac{3H}{4\pi R_A^3}+\frac{\sqrt{3\pi\lambda+4\pi^2\lambda^2R_A^2}}{2\pi G_4R_A}-\frac{\lambda}{G_4},
\label{PR}
\end{alignat}
combined with (\ref{eoso}), one can also express $T$ as a function of $R_A$:
\begin{alignat}{1}
T(R_A)=\frac{1}{2\pi R_A}+\frac{3G_4H\sqrt{3\pi\lambda+4\pi^2\lambda^2 R_A^2}}{4\pi^2\lambda R_A^3}. \label{TR}
\end{alignat}

However, it is hard to get the analytical relation $T=T(P)$ from (\ref{PR}) and (\ref{TR}),
so the JT coefficient can not be easily obtained from its definition (\ref{mu1}).
Fortunately, there is a convenient formula for the JT coefficient
\begin{alignat}{1}
\mu=\frac{1}{C_P}\left[T\left(\frac{\partial V}{\partial T}\right)_P-V\right], \label{mu2}
\end{alignat}
where $C_P$ is the heat capacity at constant $P$. From (\ref{entropy}) and (\ref{eoso}),
the heat capacity is obtained
\begin{alignat}{1}
&C_P=T\left(\frac{\partial\overset{m}{S}}{\partial T}\right)_P
=T\frac{\left(\frac{\partial\overset{m}{S}}{\partial R_A}\right)_P}{\left(\frac{\partial T}{\partial R_A}\right)_P}
\nonumber \\
=&\frac{2\pi^2 T R_A^4\sqrt{3\pi\lambda + 4\pi^2\lambda^2 R_A^2}}{G_4\sqrt{R_A^2+l^2}
[(1-2\lambda\pi R_A^2)\sqrt{3\pi\lambda+4\pi^2 \lambda^2R_A^2}+4\pi^2\lambda R_A^3(\lambda+G_4P)]},
\end{alignat}
so the JT coefficient is also obtained
\begin{alignat}{1}
\mu=&\frac{1}{C_P}\left[T\left(\frac{\partial V}{\partial T}\right)_P-V\right]
=\frac{1}{C_P}\left[T\frac{\left(\frac{\partial V}{\partial R_A}\right)_P}{\left(\frac{\partial T}{\partial R_A}\right)_P}-V\right]
\nonumber \\
=&\frac{2G_4(8\pi^2\lambda T R_A^3+9\pi T R_A-2\pi\lambda R_A^2-3)\sqrt{R_A^2+l^2}}{3\pi T R_A(3+4\pi\lambda R_A^2)}.
\end{alignat}
It has an inversion temperature at
\begin{alignat}{1}
T_i=\frac{2\pi\lambda R_A^2+3}{\pi R_A(8\pi\lambda R_A^2+9)}, \label{it}
\end{alignat}
combined with (\ref{eoso}), one can get the inversion pressure
\begin{alignat}{1}
P_{i}=\frac{4(\pi\lambda R_A^2+1)\sqrt{4\pi^2\lambda^2R_A^2+3\pi\lambda}}{\pi G_4R_A(8\pi\lambda R_A^2+9)}
-\frac{\lambda}{G_4}.
\end{alignat}
The existence of inversion point means that the FRW universe can not be always cooling or heating in the brane world scenario.

\newpage

\begin{figure}[h]
\includegraphics[height=5cm]{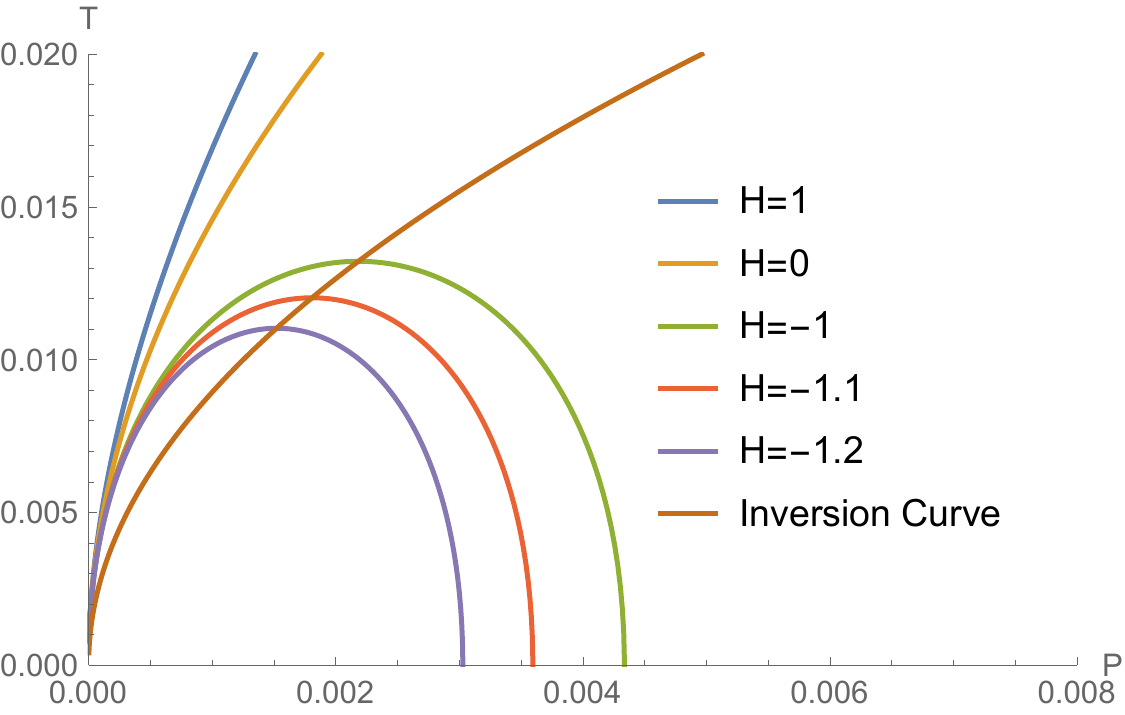}
\includegraphics[height=5cm]{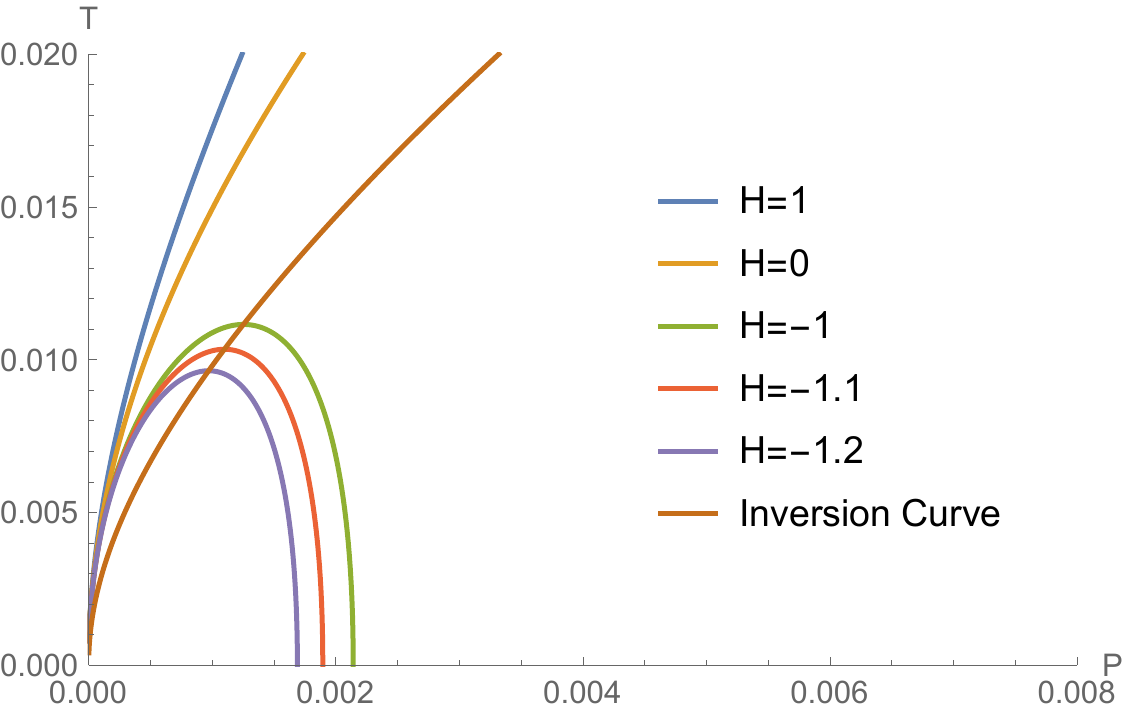}
\caption{Isenthalpic lines. In the left figure, we choose $\lambda=1$. In the right figure, we choose $\lambda=0.01$. In both figures,
we set $G_4=1$.}\label{FIG.2}
\end{figure}

From (\ref{PR}) and (\ref{TR}), we draw some isenthalpic (equal-enthalpy) lines by choosing some specific values of
$\lambda$ and $H$, which shows the existence of inversion points, see FIG. \ref{FIG.2}.

From the figure, one can see that in the brane world scenario, if $H$ is positive or zero,
there is no inversion point in the Joule-Thomson expansion. However, if $H$ is negative, there are inversion points.
The inversion point divides the Joule-Thomson expansion into the cooling and heating stages,
and the inversion curve separates cooling and heating regions. One can also see that the lower the enthalpy $H$ is,
the lower the inversion point is. One can also see that the inversion points are affected by the brane tension,
i.e. the smaller the brane tension $\lambda$ is, the lower the inversion point is. This provides a way to test the brane tension.

\section{Conclusions and Discussions}

In this paper, we have investigated the thermodynamic properties of the FRW universe in the brane world scenario,
where we have taken the RSII model as an example.
From the first law of thermodynamics (\ref{flo}) at the apparent horizon of the FRW universe,
we find that the work density $\overset{m}{W}$ can be identified with the thermodynamic pressure $P$.
We have constructed the equation of state (\ref{eoso}) for the thermodynamic pressure,
but it has no corresponding $P$-$V$ (or ``small-large") phase transition.
We have also investigated the Joule-Thomson expansion in the brane world scenario,
and got the JT coefficient. There are inversion points in the Joule-Thomson expansion,
and the values of the inversion temperature and pressure are affected by the brane tension, which provide a new way to
test the brane world scenario or extra dimension.

The results in this paper shows that if one defines the thermodynamic pressure $P$ of the FRW universe
from the work density, one can get the standard first law of thermodynamics and  construct an equation of state $P=P(V,T)$,
which is very essential for further study of many thermodynamic behaviors. One can also study the thermodynamics of the FRW universe
in other brane world models, such as the DGP model \cite{Dvali:2000hr, Dvali:2000xg, Zhang:2004in, Sheykhi:2007zp}, Gauss-Bonnet brane world \cite{Sheykhi:2007gi}, etc, where new phenomena may be found. The thermodynamic behavior of black holes in the brane world scenario is
also an interesting question.

\section*{Acknowledgments}

We are very grateful to the reviewer and editor for their valuable comments and questions. We also thank L.M. Cao, T. Jacobson, Y.X. Liu,
T.T. Sui, Y.Q. Wang, S.W. Wei, Z.M. Xu, J. Yang, Y.H. Yin, et.al. for helpful discussions. S.B.Kong, H. Abdusattar and Y.P. Hu are supported by National Natural Science Foundation of China (No.12175105). H.Zhang is supported by National Natural Science Foundation of China (No.12235019)
and the National Key Research and Development Program of China (No. 2020YFC2201400).

\appendix

\section{FRW Universe in Physical Coordinate System}

One can rewrite the line-element (\ref{metric}) of the FRW universe by physical coordinate system $(t,R,\theta,\varphi)$ as
\cite{Cai:2008gw}
\begin{alignat}{1}
\od s^2=-\frac{1-R^2/R_A^2}{1-k R^2/a^2}\od t^2-\frac{2HR}{1-kR^2/a^2}\od t\od R
+\frac{1}{1-kR^2/a^2}\od R^2+R^2(\od\theta^2+\sin^2\theta\od\varphi^2). \label{nm}
\end{alignat}
For dynamical spacetime, there is a Kodama vector field defined as \cite{Kodama:1979vn,Cai:2008gw}
\begin{alignat}{1}
K^a:=-\epsilon^{ab}\nabla_b R,
\end{alignat}
which generates a preferred time flow (dynamic time \cite{Hayward:1997jp}) and can be treated as the dynamical version
of the Killing vector field \cite{Hayward:1998ee}.
For the new line-element (\ref{nm}) of the FRW universe, the Kodama vector has only one non-zero component
\begin{alignat}{1}
K^0=-\sqrt{1-k\frac{R^2}{a^2}},
\end{alignat}
so its norm is
\begin{alignat}{1}
K_a K^a=-\left(1-\frac{R^2}{R_A^2}\right),
\end{alignat}
which shows that the Kodama vector is timelike, spacelike, or null if $R$ is less, larger,
or equal to $R_A$ respectively.

Similar to the Killing vector, one can define conserved current
for dynamical spacetime from the Kodama vector
\begin{equation}
J^{\mu}:=-T^{\mu}_{~~\nu}K^{\nu},
\end{equation}
which is divergence free
\begin{equation}
\nabla_{\mu}J^{\mu}=0
\end{equation}
in general relativity, and many modified theories of gravity, etc.
Then, one can define the associated conserved charge \cite{Hayward:1994bu}
\begin{equation}
Q_J:=\frac{1}{8\pi G_4}\int_{\Sigma} J^{\mu}\mathrm{d}\Sigma_{\mu},
\end{equation}
where $\Sigma$ is a hypersurface and $\mathrm{d}\Sigma_{\mu}$ is its directed surface element.
For many cases \cite{Hu:2015xva,Hu:2016hpm}, the above conserved charge is just the generalized
Misner-Sharp energy/mass in the same gravitational theory,
so it is called the conserved charge method of the Misner-Sharp energy.

\section{Equation of State for $P=W$}

From (\ref{fe1}) and (\ref{fe2}), one can get the work density defined by the total energy-momentum tensor
\begin{alignat}{1}
W=-\frac{1}{2}h_{ab}T^{ab}=\frac{1}{2}(\rho+\rho_e-p-p_e)=\frac{3}{8\pi G_4}\left(H^2+\frac{k}{a^2}\right)
+\frac{1}{8\pi G_4}\left(\dot{H}-\frac{k}{a^2}\right),
\end{alignat}
combined with (\ref{ra}), (\ref{dotra}), and (\ref{kappa}), one can get
\begin{alignat}{1}
W=-\frac{\kappa}{4\pi G_4 R_A}+\frac{1}{8\pi G_4R_A^2}.
\end{alignat}
Together with (\ref{KHT}) and (\ref{tp1}), one can get the equation of state
\begin{alignat}{1}
P(R_A,T)=\frac{T}{2 G_4 R_A}+\frac{1}{8\pi G_4R_A^2}, \label{eos}
\end{alignat}
which resembles the one in GR \cite{Abdusattar:2021wfv,Kong:2021dqd} as expected. This equation shows no $P$-$V$ phase transition.

{}

\end{document}